\documentclass[a4paper]{jpconf}
\usepackage{graphicx}
\begin{document}

\def\bfg #1{{\mbox{\boldmath $#1$}}}

\title{  Spin-dependent
$\bar p d$ cross sections at low and intermediate energies}

%



\author{Yu~N~Uzikov$^1$ and J~Haidenbauer$^2$
}



\address{
$^1$Laboratory of Nuclear Problems, Joint Institute for Nuclear
Research, 141980 Dubna, Russia\\
$^2$Institute for Advanced Simulation and J\"ulich Center for Hadron Physics,
Forschungszentrum
$\phantom{^1}$J\"ulich, D-52425 J\"ulich, Germany}
\ead{uzikov@nusun.jinr.ru}


\begin{abstract}
Antiproton-deuteron (${\bar p}d$) scattering is calculated at beam energies
below 300 MeV within the Glauber approach, utilizing the amplitudes of the
J\"ulich $\bar N N$ models. A good agreement is obtained
with available experimental data on upolarized  differential
and integrated $\bar p d$ cross sections.
Predictions for polarized total ${\bar p}d$ cross
sections are presented, obtained within the single scattering approximation 
including Coulomb-nuclear interference effects.
It is found that the total longitudinal and transversal ${\bar p }d$
cross sections are comparable in absolute value to those for
${\bar p}p$ scattering.
 The kinetics of polarization buildup is considered.
%
\end{abstract}

\section{Introduction}
\label{intro}

The preparation of an intense beam of polarized antiprotons is
{\it the} crucial point for the physics program proposed by the PAX 
collaboration \cite{PAX} at the future FAIR facility in Darmstadt.
A possibility to overcome this experimental challenge is seen in elastic
scattering of antiprotons off a polarized $^1$H target \cite{Rathmann}.
This conjecture is
motivated by the result of the FILTEX experiment \cite{FILTEX},
where a sizeable effect of polarization buildup was achieved in a
storage ring by scattering of unpolarized protons off polarized hydrogen
atoms at low beam energies of 23 MeV.
Recent theoretical analyses \cite{MS,NNNP,NNNP1} suggest that
the polarization effect observed in Ref.~\cite{FILTEX} is
solely due to the spin dependence of the hadronic (proton-proton)
interaction, which gives rise to the so-called spin-filtering mechanism,
i.e. leads to different rates of removal of beam protons from the ring for
different polarization states of the
hydrogen target.
Contrary to what was assumed before \cite {HOM}, proton scattering on the
polarized electrons of hydrogen atoms does not provide sizeable effects
for the polarization buildup \cite{MS,NNNP}.
Accordingly, only the hadronic interaction of antiprotons with
nucleons or nuclei can be used to produce polarized
antiprotons on the basis of the spin-filtering mechanism \cite{MS}.
Since the spin-dependent part of the $\bar pN$
interaction is still poorly known experimentally,
the polarization buildup mechanism in elastic scattering of stored
antiprotons off a polarized $^1$H target is planned to be studied
in a new experiment at CERN \cite{AD,AD1} at intermediate energies.
Some theoretical estimations of the expected polarization effects
were already presented, based on the amplitudes of the 
Paris~\cite{DmitrievMS} and J\"ulich \cite{own,mine} $\bar NN$ potential
models and the Nijmegen \cite{DMS2} $\bar NN$ partial-wave analysis.

In this context, it is important to explore other 
antiproton--nucleus interactions 
as possible source for the antiproton polarization buildup too.
Therefore, we present here results of a study of polarization effects
in antiproton-deuteron (${\bar p} d$) scattering for beam energies
up to 300 MeV \cite{own}. Besides the issue of polarization buildup for
antiprotons, $\bar{p}$ scattering on a polarized deuteron,
if it will be studied experimentally, can be also used
as a test for our present knowledge of the ${\bar p}n$ and
${\bar p}p$ interactions.
Our investigation is based on the Glauber-Sitenko theory for
${\bar p}d$ scattering and it utilizes the ${\bar N}N$ interaction models
developed by the J\"ulich group \cite{Hippchen,Mull}
as input for the elementary amplitudes.

\section{Spin dependence of the total $\bar p d$ cross section}

Considering the full spin dependence of the forward $\bar p d$ elastic scattering
amplitude \cite{rekalo,own} and using the optical theorem, one can
show that the total polarized ${\bar p} d$ cross section can be written as
\begin{equation}
\label{totalspin}
\sigma_{tot}=\sigma_0+\sigma_1{\bf P}^{\bar p}\cdot {\bf P}^d+
 \sigma_2 ({\bf P}^{\bar p}\cdot {\bf m}) ({\bf P}^d\cdot {\bf m})+
\sigma_3 P_{zz},
\end{equation}
 where ${\bf P}^{\bar p}$ is the polarization of the
 antiproton beam
 and ${\bf P}^d$ ($P_{zz}$) is the vector (tensor) polarization
 of the 
deuterium target, $\sigma_0$ is the unpolarized and $\sigma_i$ ($i=1,2,3$) are
 the polarized total cross sections. The unit vector ${\bf m}$ is fixed by the
 direction of the beam momentum.
 One can find from Eq.~(\ref{totalspin}) that only the cross sections
 $\sigma_1$ and $\sigma_2$ are connected with the spin-filtering mechanism \cite{own}
 and, thus, determine the rate of the polarization buildup in the scattering of
 unpolarized antiprotons off polarized deuterons.
 The tensor cross section $\sigma_3$ is not related to the
 polarization of the beam and, therefore, is not relevant
 for the spin-filtering. However, this cross section, as well as the
 unpolarized cross section $\sigma_0$, determine the lifetime of the beam.

We utilize the Glauber theory of multiple scattering \cite{FrancoGlauber}
for investigating the $\bar p d$ scattering process.
For the elementary $\bar p N$ amplitudes we use those of the
J\"ulich models A and D \cite{Hippchen,Mull}.
Details on the applied formalism can be found in Ref.~\cite{own}.
In order to check the reliability of this approach we
calculate the unpolarized total and differential $\bar p d$
cross sections where we can compare our results with available
experimental information.

In the evaluation of the polarized cross sections $\sigma_i$
($i=1,2,3)$, we take into account the Coulomb-nuclear interference terms,
which are added to the corresponding purely hadronic
cross sections. The pure Coulomb amplitude does not contribute to
$\sigma_i, \ i=1, 2, 3$, but it gives an important contribution to
$\sigma_0$. In order to calculate the contribution
of the Coulomb-nuclear interference terms one cannot use the optical
theorem because of the Coulomb singularity at the scattering angle
$\theta = 0^\circ$, and therefore we use here the method of
Ref.~\cite{MS}, adapted for the case of $\bar pd$ scattering
\cite{own}. For the polarized cross sections we
use only the single-scattering approximation.

\section{Results and discussion of $\bar p d$ scattering}

As was shown in Refs.~\cite{Kondratyuksb,Dalkarov},
in forward elastic scattering of antiprotons off
nuclei the Glauber theory of diffractive multiple scattering,
though in principle a high-energy approach,
works rather well even at fairly low antiproton beam energies.
The reason for this is that due to strong annihilation effects,
the $\bar p N$ elastic differential cross section
is peaked in forward direction already at rather low energies
and, therefore, suitable for application of the
eikonal approximation, which is the basis of the Glauber theory.

The elastic spin-averaged $\bar p N$ scattering amplitude can be
 parameterized as
\begin{equation}
\label{fpn}
  f_{\bar pN}(q)=\frac{k_{\bar pN}\sigma^{\bar p N}_{tot}(i+\alpha_{\bar p N})}
{4\pi} \exp{(-\beta^2_{\bar p N}q^2/2)},
 \end{equation}
 where $\sigma^{\bar p N}_{tot}$ is the total unpolarized $\bar p N$
 cross section, $\alpha_{\bar p N}$ is the ratio of the real to
 imaginary part of the forward amplitude $f_{\bar pN}(0)$,
 $\beta_{\bar p N}^2$ is the slope of the diffraction cone, $q$ is the
 transferred 3-momentum, and $k_{\bar p N}$ is the $\bar p N$ cms momentum.
We use Eq.~(\ref{fpn}) to represent the scattering amplitudes of the J\"ulich
$\bar NN$ models in analytical form.
When performing the fit we found that even at beam energies as low as 10-25 MeV
the parameter $\beta^2_{\bar p N}$ is large, i.e. $40 - 50$ (GeV/c)$^2$,
reflecting the fact that the $\bar p N$ amplitude is indeed peaked in forward
direction.

\begin{figure}
 \includegraphics[width=0.65\textwidth,angle=-90]{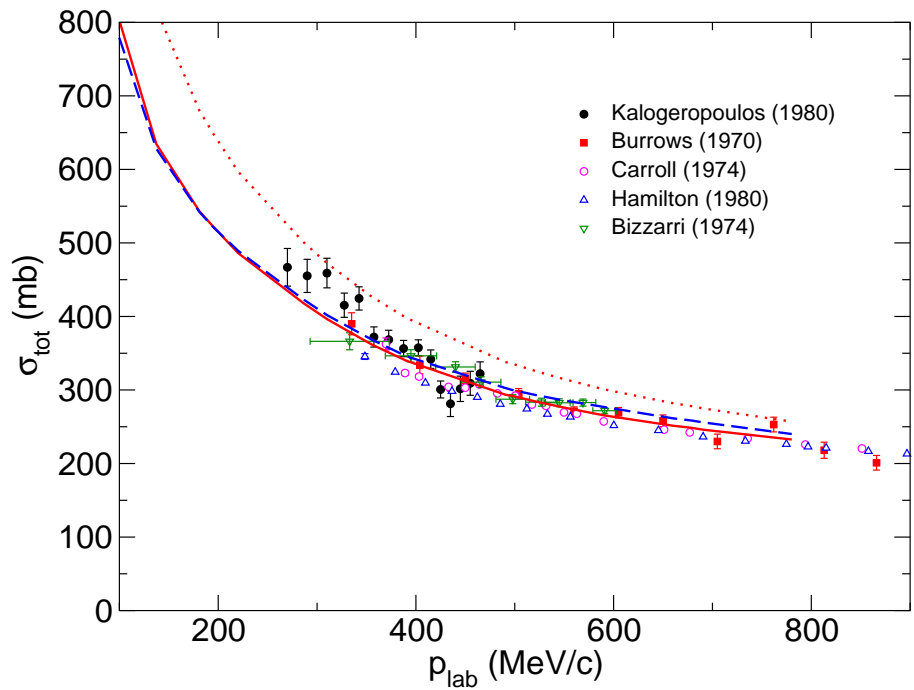}
\caption{Total $\bar p d$ cross section versus the beam momentum
$p_{lab}$. The solid and dashed lines are results based on the
$\bar NN$ models D and A, respectively. The dotted line is the
results for model D obtained within the single-scattering approximation.
Data are taken from Refs.
\cite{BizzarriNC74,Kalogeropoulos,Burrows,Carroll,Hamilton}.
}
\label{totpd}
\end{figure}

Results for the total unpolarized $\bar p d$ cross
section are displayed in Fig.~\ref{totpd} together with experimental
information \cite{BizzarriNC74,Kalogeropoulos,Burrows,Carroll,Hamilton}.
One can see that the single-scattering approximation (shown here
for model D only) overestimates the total unpolarized cross section
by roughly 15\%, cf. the dotted line.
But the shadowing effect generated by $\bar p N$ double scattering
reduces the cross section (solid line) and leads to a good agreement with
the experiment. The results for model A (including also double scattering)
are very similar (dashed line) and also in agreement with the data.

\begin{figure}
\vspace{1.5cm}
 \includegraphics{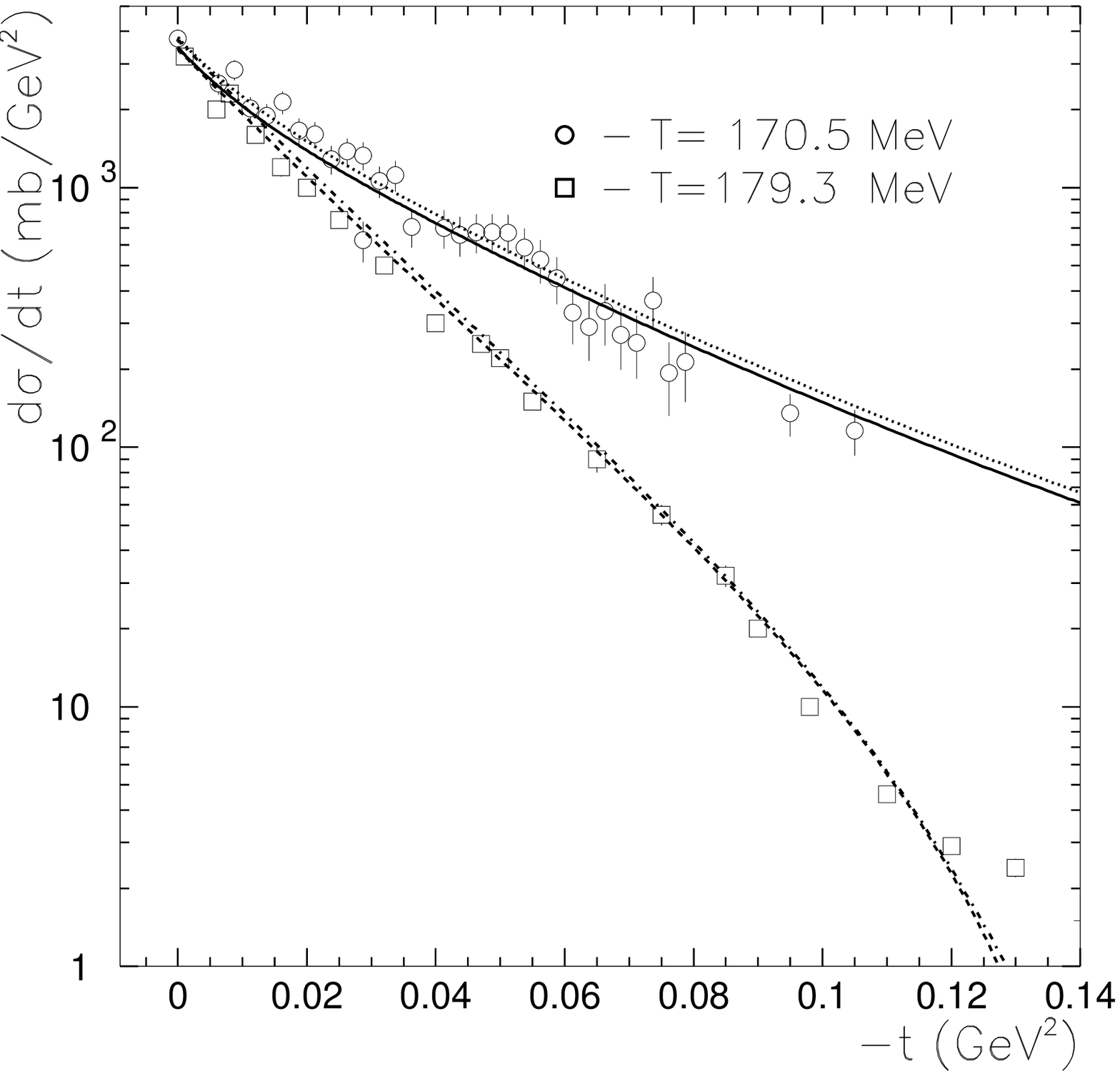}
 \includegraphics{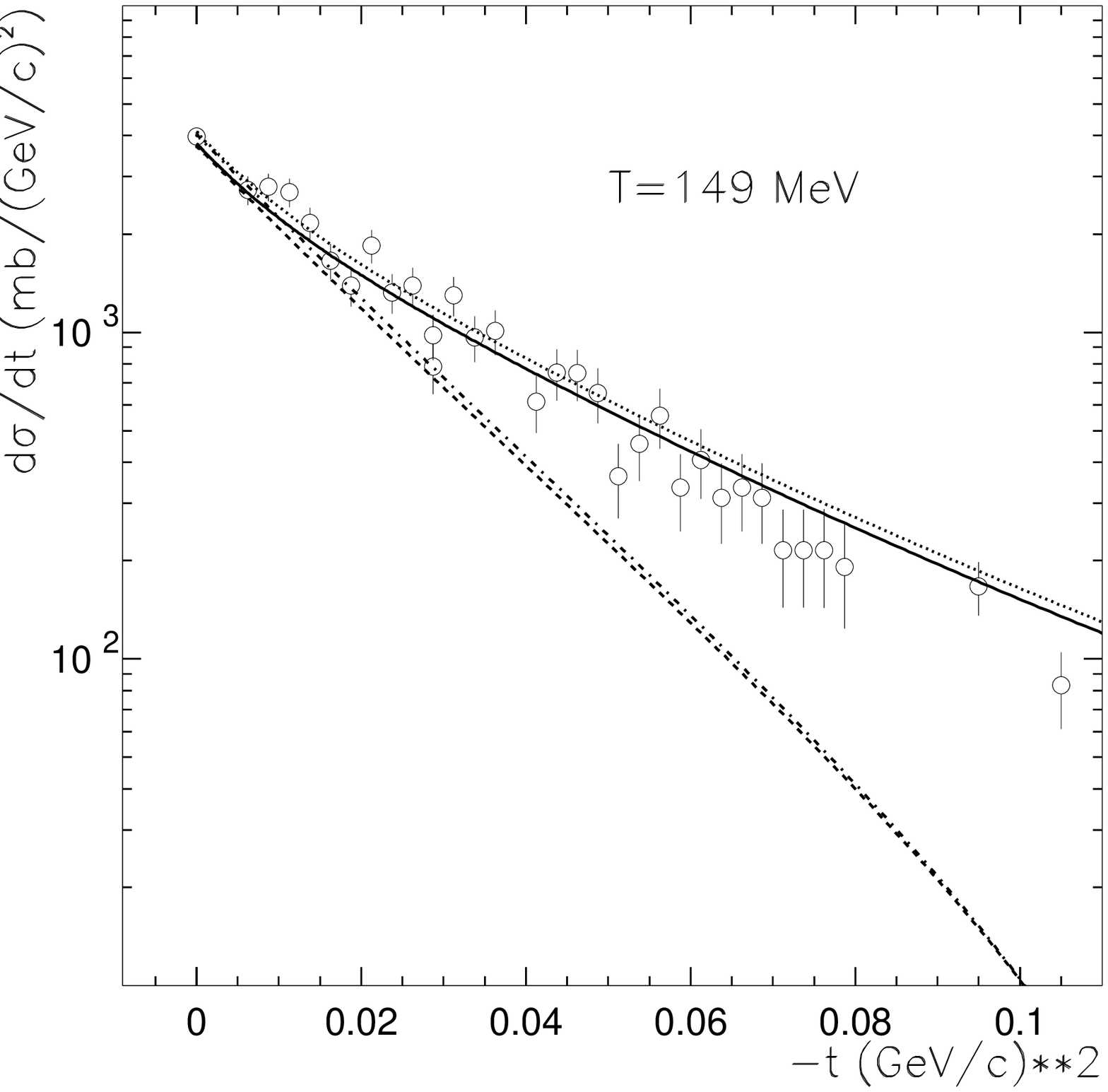}
 \includegraphics{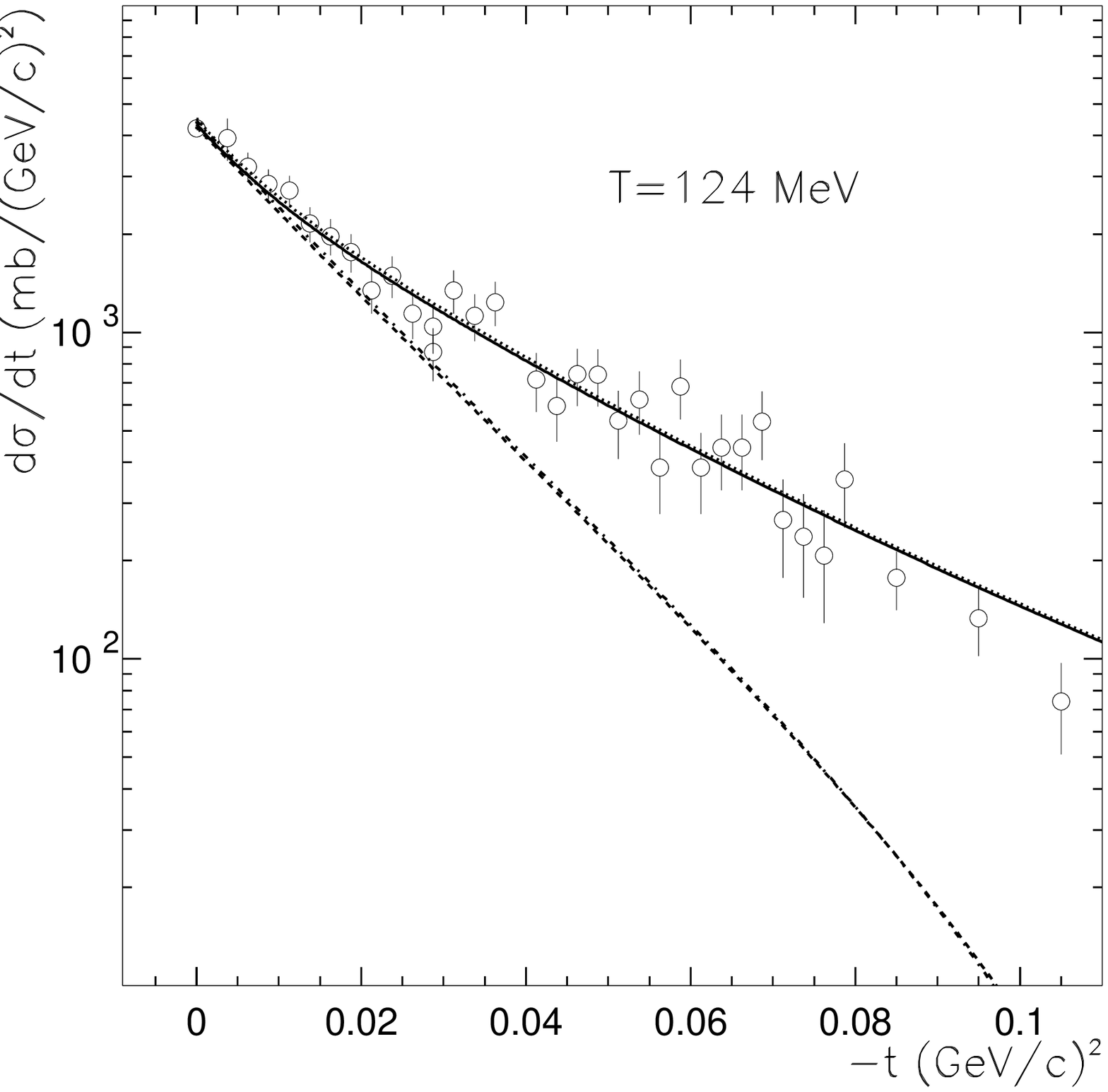}
 \includegraphics{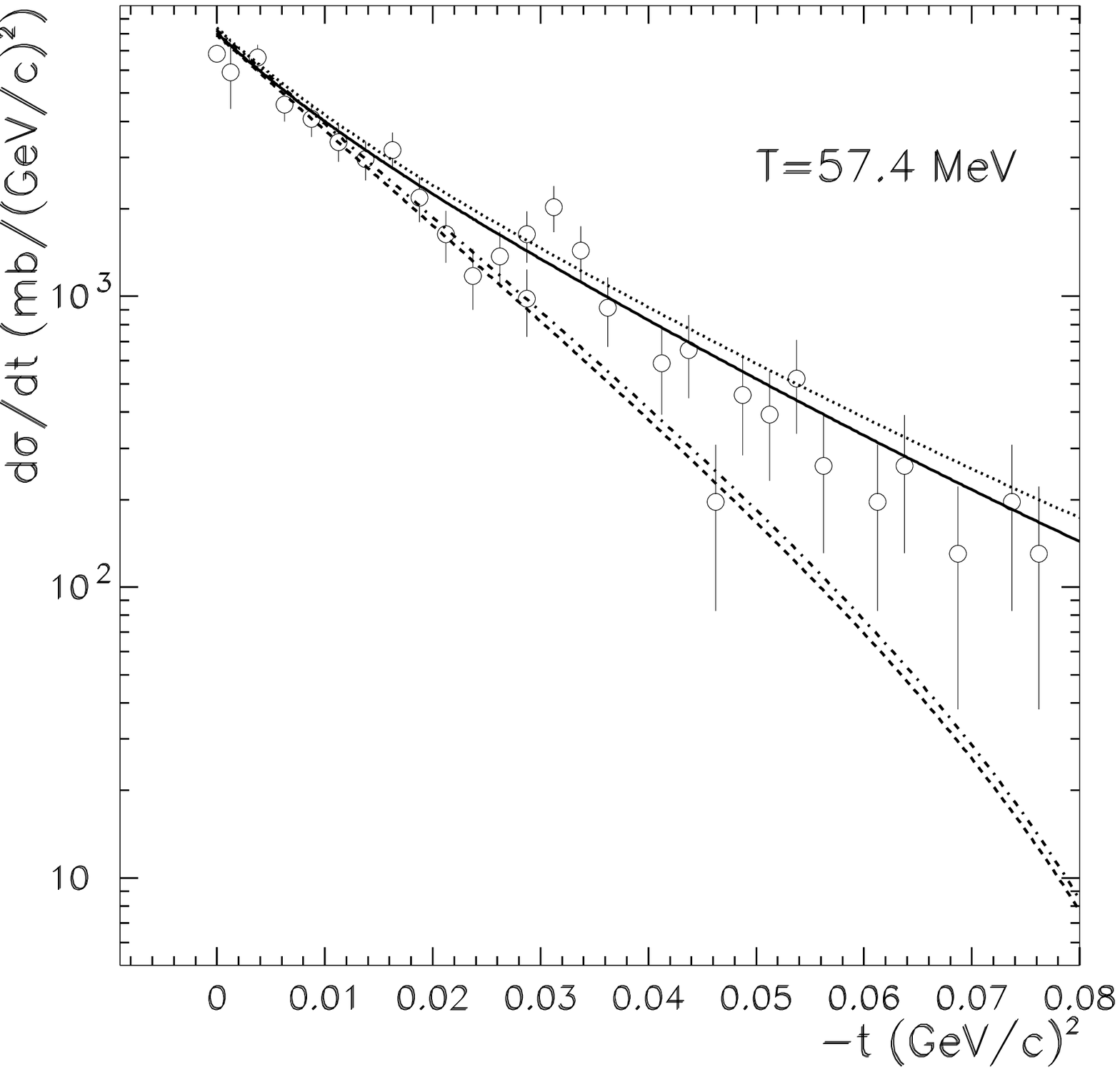}
 \vspace{11.7cm}

\caption{Elastic (lower lines) and elastic plus inelastic (upper lines)
 $\bar p d$ differential cross sections versus the transferred
 momentum for different antiproton beam energies.
 The lines are results of a calculation based on the Glauber theory
 for model A (dotted and dashed-dotted) and D (solid and dashed)
 utilizing the parameterizations of the $\bar pN$ amplitudes via
 Eq.~(\ref{fpn}) as given in Ref.~\cite{own}.
Data are taken from Ref. \protect\cite{bruge88} 
(179.3 MeV, squares)
and from Ref. \cite{BizzarriNC74} (57.4--170.5 MeV, circles).
}
\label{pbard179}
\end{figure}

Predictions for differential cross sections are presented in
Fig.~\ref{pbard179}.
Also here  the single-scattering mechanism as well as the
double-scattering terms were included in the corresponding calculation.
The ABB form factor \cite{ABB} is used for the deuteron.
At $T_{lab}$ = 179.3 MeV data for the elastic differential
cross section are available \cite{bruge88}. These data
(squares in Fig.~\ref{pbard179}) are nicely reproduced by
our model calculation for forward angles. Also the
differential cross sections for elastic (${\bar p}d\to {\bar p} d$)
plus inelastic (${\bar p}d\to {\bar p} pn$) scattering events,
measured at the neighboring energy of $T_{lab}$ = 170 MeV as well
as at some lower energies \cite{BizzarriNC74} (circles), are well
described.

\begin{figure}
\includegraphics{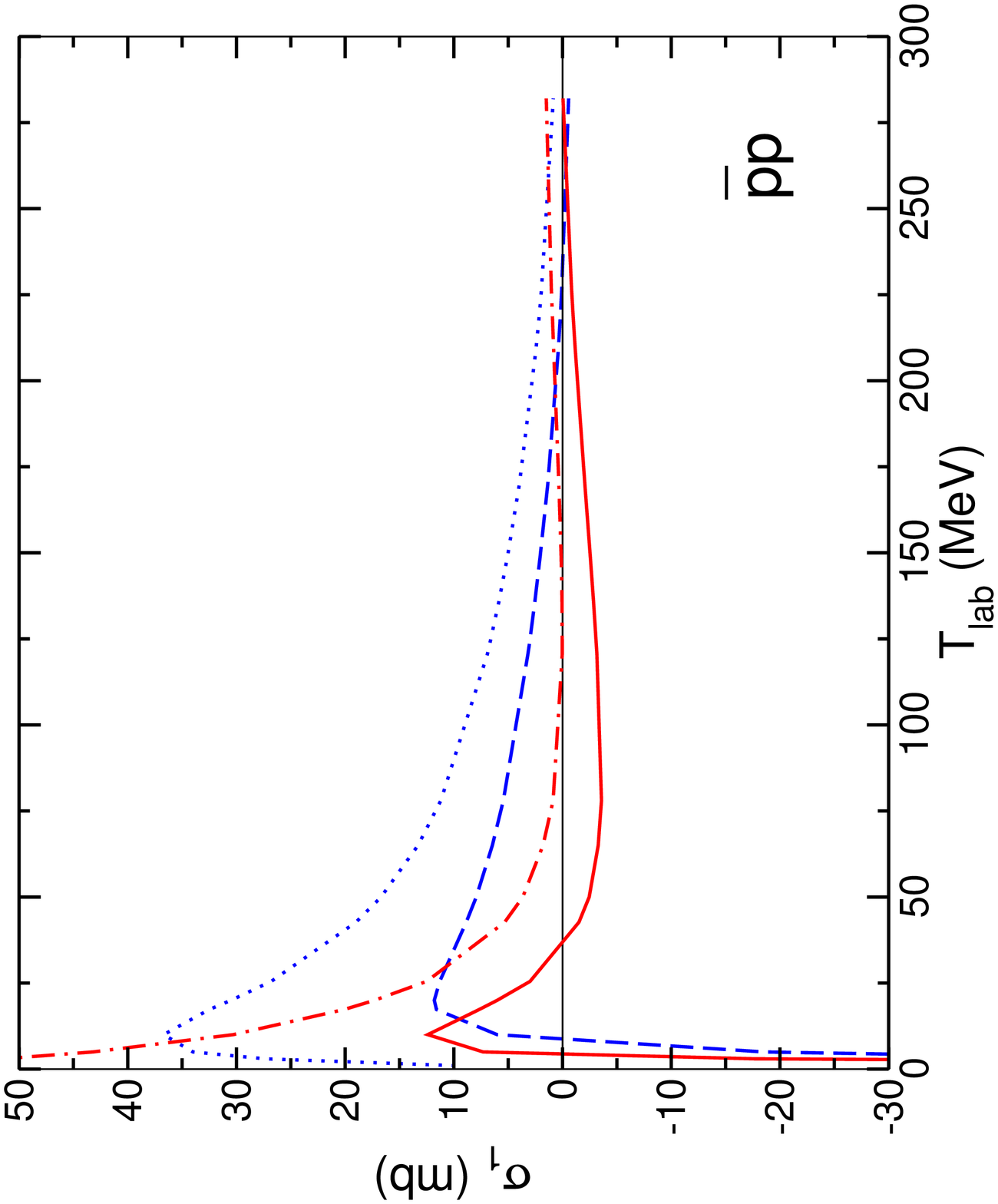}
\includegraphics{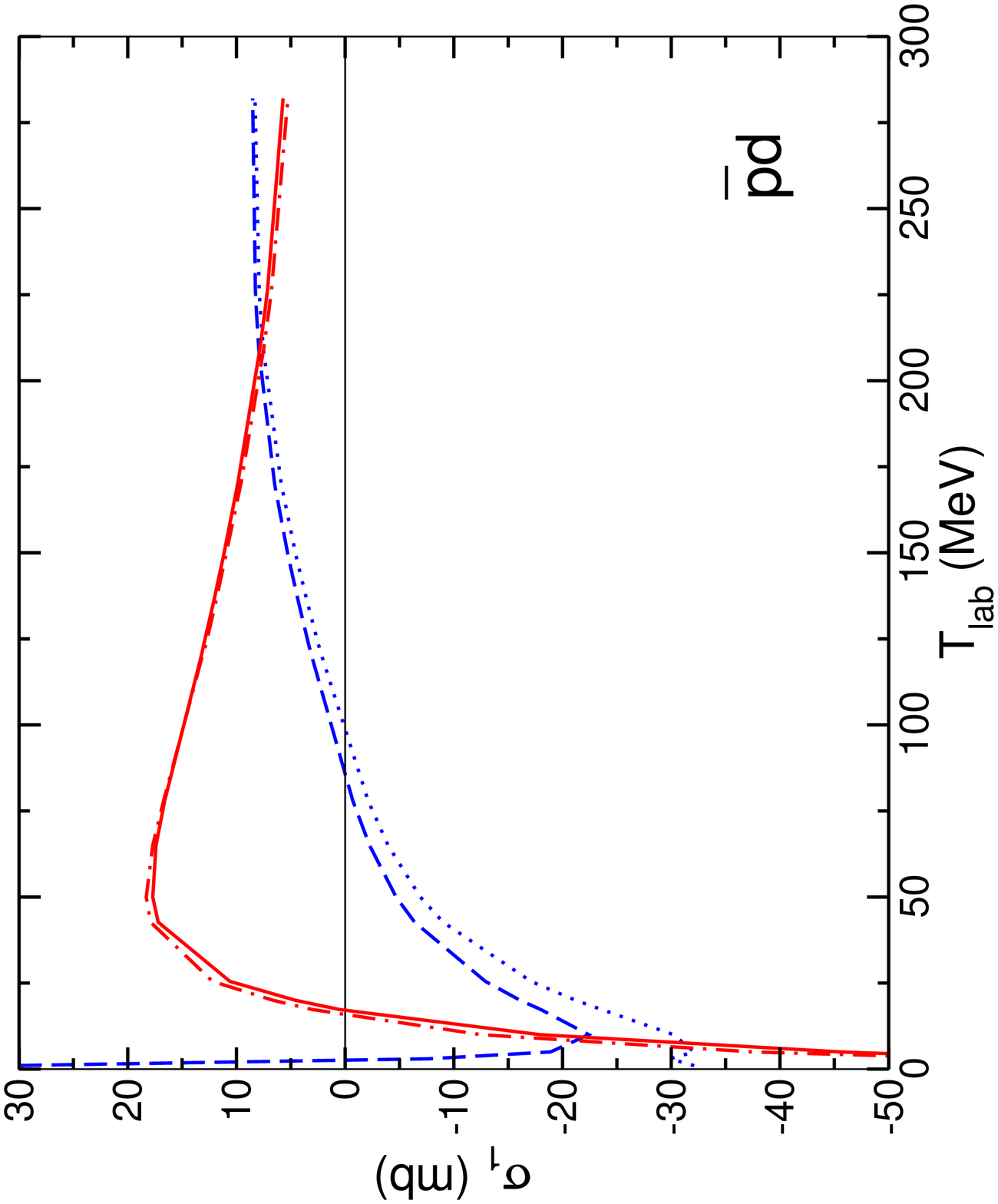}
\includegraphics{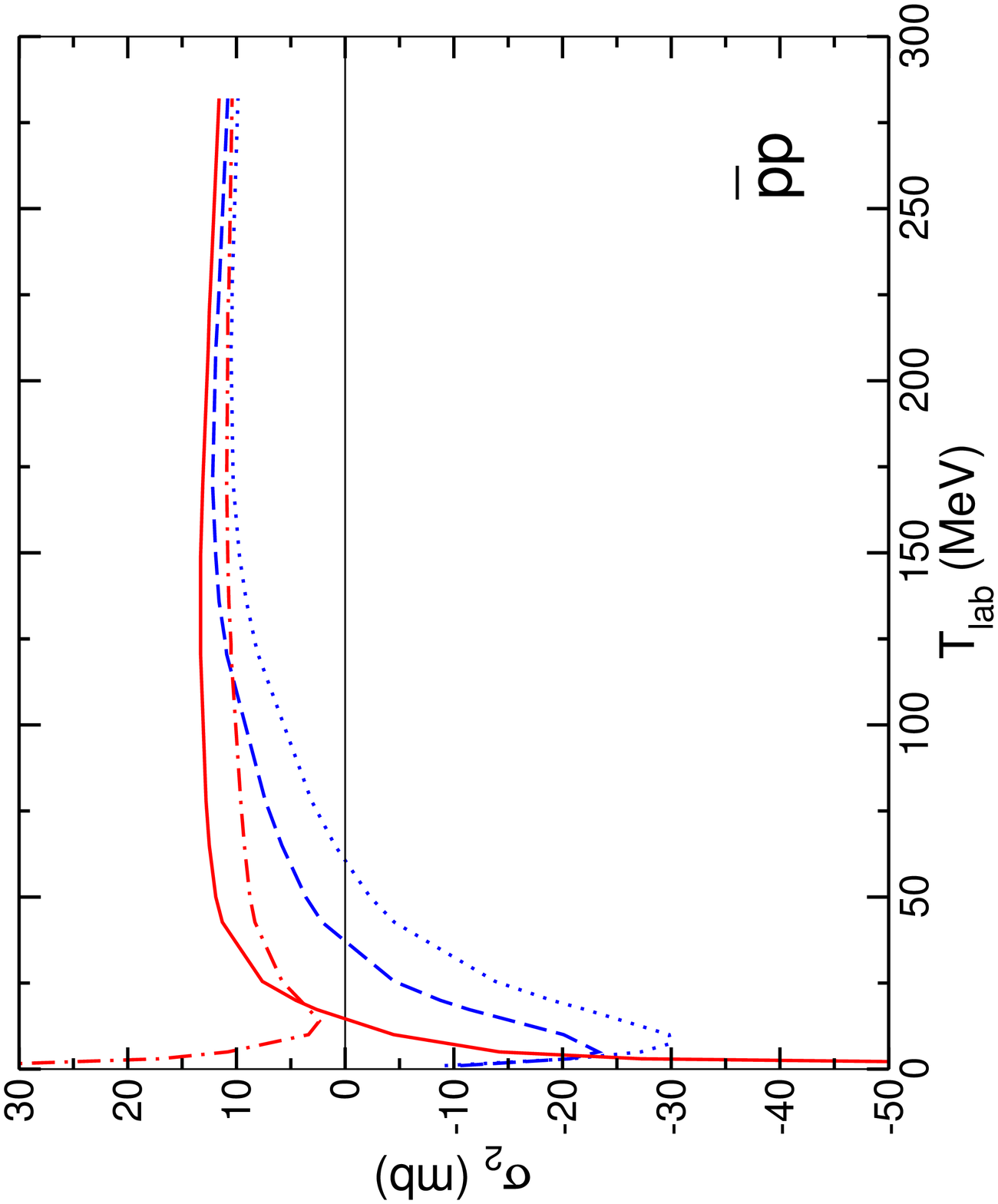}
\includegraphics{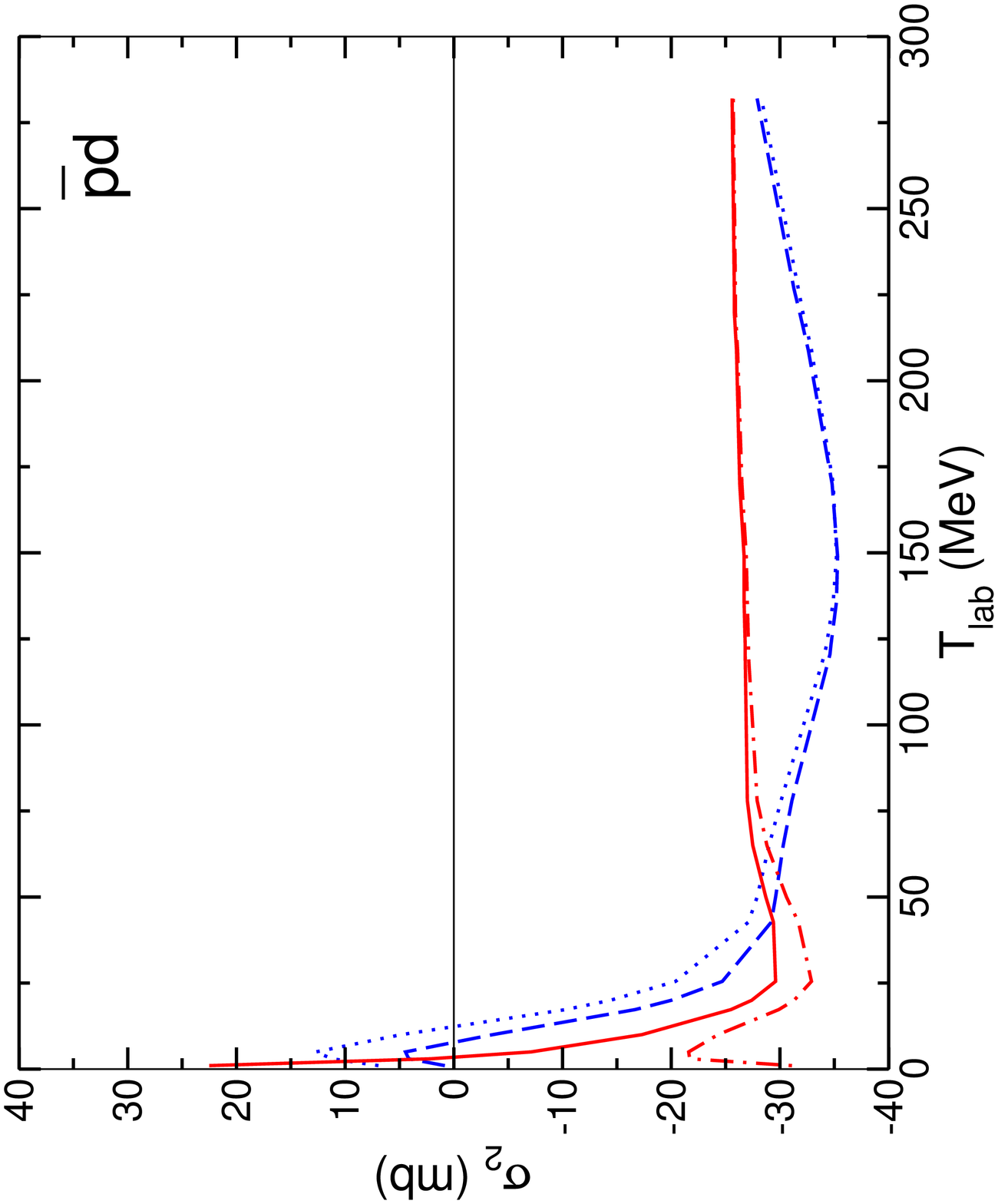}
\vskip 10.5cm
\caption{Total spin-dependent cross sections $\sigma_1$
and $\sigma_2$ versus antiproton laboratory energy $T_{lab}$
for $\bar p p$ (left column) and $\bar p d$ (right column)
scattering. Results based on the purely hadronic amplitude,
$\sigma^h_i$, (model D: dash-dotted line, model A: dotted line)
and including the Coulomb-nuclear interference term, i.e.
$\sigma_i = \sigma^h_i + \sigma^{int}_i$,
(D: solid line, A: dashed line), are shown.
}
\label{totpdx}
\end{figure}

Results for the spin-dependent $\bar p d$ cross sections $\sigma_1$ and
$\sigma_2$ are obtained in the single-scattering approximation and presented
in Fig.~\ref{totpdx} (right-hand side).
We show predictions based on the purely hadronic
part, $\sigma_i^h$, as well as full results, including
the Coulomb-nuclear interference term,
i.e. $\sigma_i=\sigma_i^h+\sigma_i^{int}$ for $i=1,2$.
The $\bar NN$ model D predicts large values for $\sigma_1$ around
40 MeV and for $\sigma_2$ around 25 MeV. In case of
model A the most pronounced spin dependence is seen at
considerably higher energies.

Compared with the results for the $\bar pp$ reaction,
shown on the left-hand side of Fig.~\ref{totpdx}, the spin-dependent
${\bar p }d$ cross sections $\sigma_1$ and $\sigma_2$ are of similar
magnitude or even larger.
With regard to the purely hadronic contribution, $\sigma_i^h$ (i=1,2), 
the cross sections 
for $\bar p p$ and $\bar p d$ are, in general, of opposite sign \cite{own}.
It should be said, however, that the sign does not affect the spin-filtering mechanism.
The effect of the Coulomb-nuclear interference is somewhat smaller
for $\bar pd$ scattering than for the $\bar pp$ case. This difference
comes from the additional $\bar p n$ amplitudes entering the expression
for $\sigma_i^{int}$ in case of the $\bar p d$ reaction, cf.
Ref.~\cite{own} for details.

One can see from Fig.~\ref{totpdx} that the largest values for the
polarized $\bar p d$ cross sections (and also those for $\bar p p$)
are expected at very low
energies, i.e. for $T_{lab}$ less than 10 MeV, where the cross sections
are dominated by the Coulomb-nuclear interference term.
However, as was already mentioned above, at these energies the pure Coulomb cross section
becomes rather large, so that the method of spin-filtering for the polarization buildup
cannot be applied due to the decrease of the beam lifetime.

Acording to the analysis of the kinetics of polarization \cite{MS,NNNP1},
the polarization buildup is determined mainly by the ratio of the polarized
total cross sections to the unpolarized one ($\sigma_0$)
\cite{MS}. Let as define the unit vector ${\bfg \zeta}= {\bf P}_T/ P_T$,
 where  $ {\bf P}_T={\bf P}^d$ is the  target polarization vector, which in the
 case of $\bar p d$ scattering  enters 
 Eq. (\ref{totalspin}). The non-zero antiproton beam polarization vector
${\bf P}^{\bar p}$, produced  by the polarization buildup,
 is collinear to the vector ${\bfg \zeta}$ for any directions of ${\bf P}_T$
 and can be calculated from consideration of the kinetics of polarization.
 The general solution for the kinetic
 equation for $\bar p p$ scattering is given in Ref.~\cite{MS}. Here we
 assume that this solution is valid for the $\bar p d$ scattering also.
 Therefore, for the spin-filtering mechanism of the polarization buildup
 the polarization degree at the time $t$ is given
 by \cite{MS,DMS2}
\begin{equation}
P_{\bar p}(t)=\tanh\left [\frac {t}{2}(\Omega_{-}^{out}-
\Omega_{+}^{out})\right ],
\label{pdeg}
\end{equation}
where
\begin{equation}
\Omega_{\pm}^{out}=nf\left \{\sigma_0\pm P_T\left [\sigma_1 +
({\bfg \zeta}\cdot {\bf m})^2\sigma_2\right ]\right \}.
\label{omega}
\end{equation}
Here $n$ is the areal density of the target and $f$ is the beam revolving
frequency. One should note that the  tensor cross section $\sigma_3$
from Eq. (\ref{totalspin}) does not contribute to $\Omega_{\pm}^{out}$.
Assuming the condition $|\Omega_{-}^{out}-\Omega_{+}^{out}|
<< (\Omega_{-}^{out}+\Omega_{+}^{out}$), which was found in
Refs.~\cite{MS,DMS2} for the $\bar p p$ scattering in rings at $n=10^{14}$
cm$^{-2}$ and $f=10^6$ c$^{-1}$,
one can simplify Eq. (\ref{pdeg}).
If one denotes the number of antiprotons in the beam at the time moment $t$
as $N(t)$, then the figure of merit (FOM) is $P_{\bar p}^2(t)N(t)$. This value
is maximal at the moment $t_0=2\tau$, where $\tau$ is the beam life time,
which is determined by the total cross section $\sigma_0$ of
the interaction of the antiprotons with the deuteron target as
\begin{equation}
\tau=\frac{1}{nf\sigma_0}.
\label{tau}
\end{equation}
To estimate the efficiency of the polarization buildup mechanism it is
 instructive to calculate the polarization degree $P_{\bar p}$  at the time $t_0$
 \cite{DMS2}. In our definition for $\sigma_1$ and $\sigma_2$, which
 differ from that in Refs.~\cite{MS,DMS2},
 we find
\begin{eqnarray}
P_{\bar p} (t_0)=-2P_T\frac{\sigma_1}{\sigma_0}, \,\,\, if \,\,\,
{\bfg \zeta}\cdot {\bf m}=0,
\nonumber \\
P_{\bar p} (t_0)=-2P_T\frac{\sigma_1+\sigma_2}{\sigma_0}, \,\,if\,\,\,
 |{\bfg \zeta}\cdot {\bf m}|=1,
\end{eqnarray}

\begin{figure}
\includegraphics{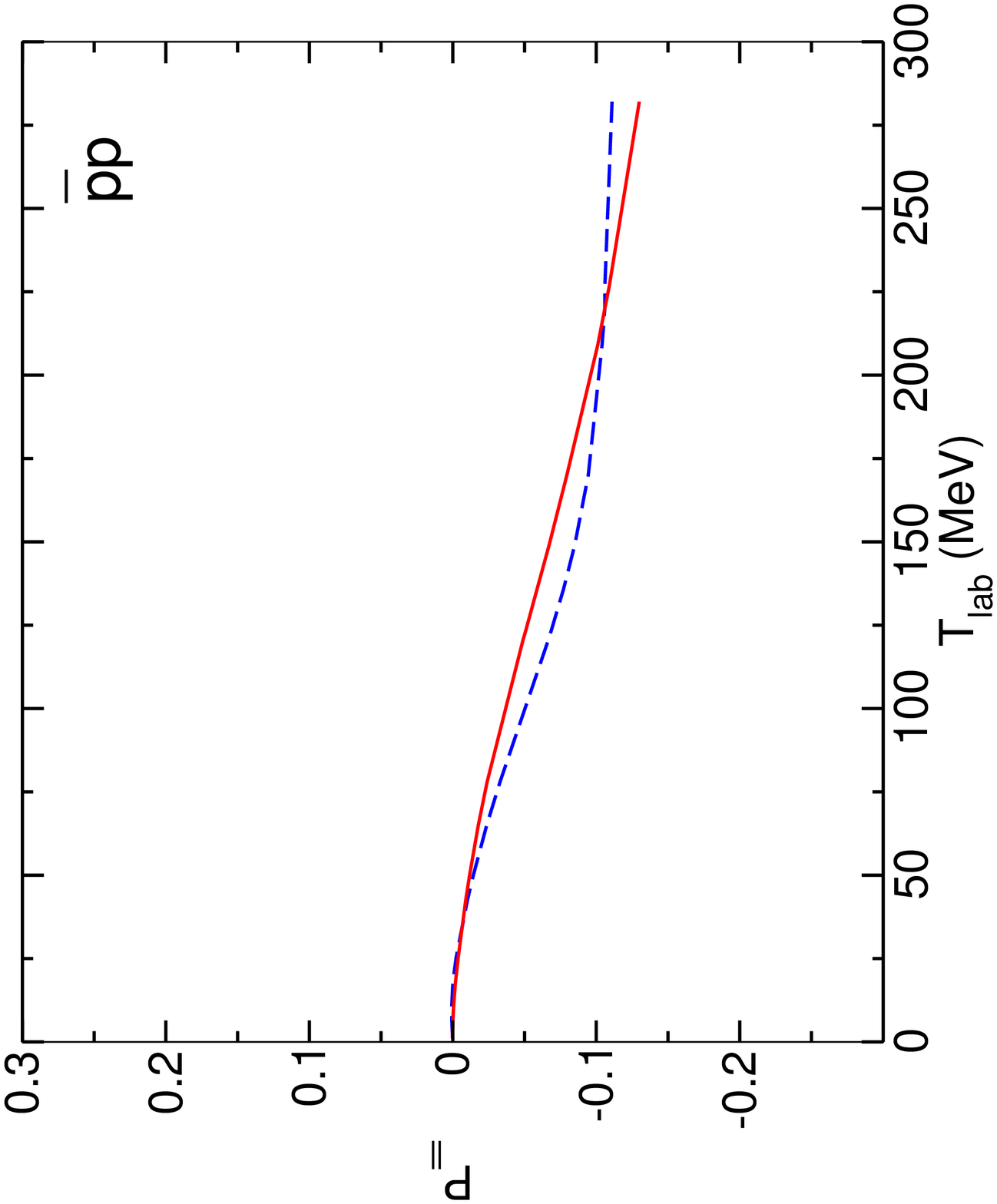}
\includegraphics{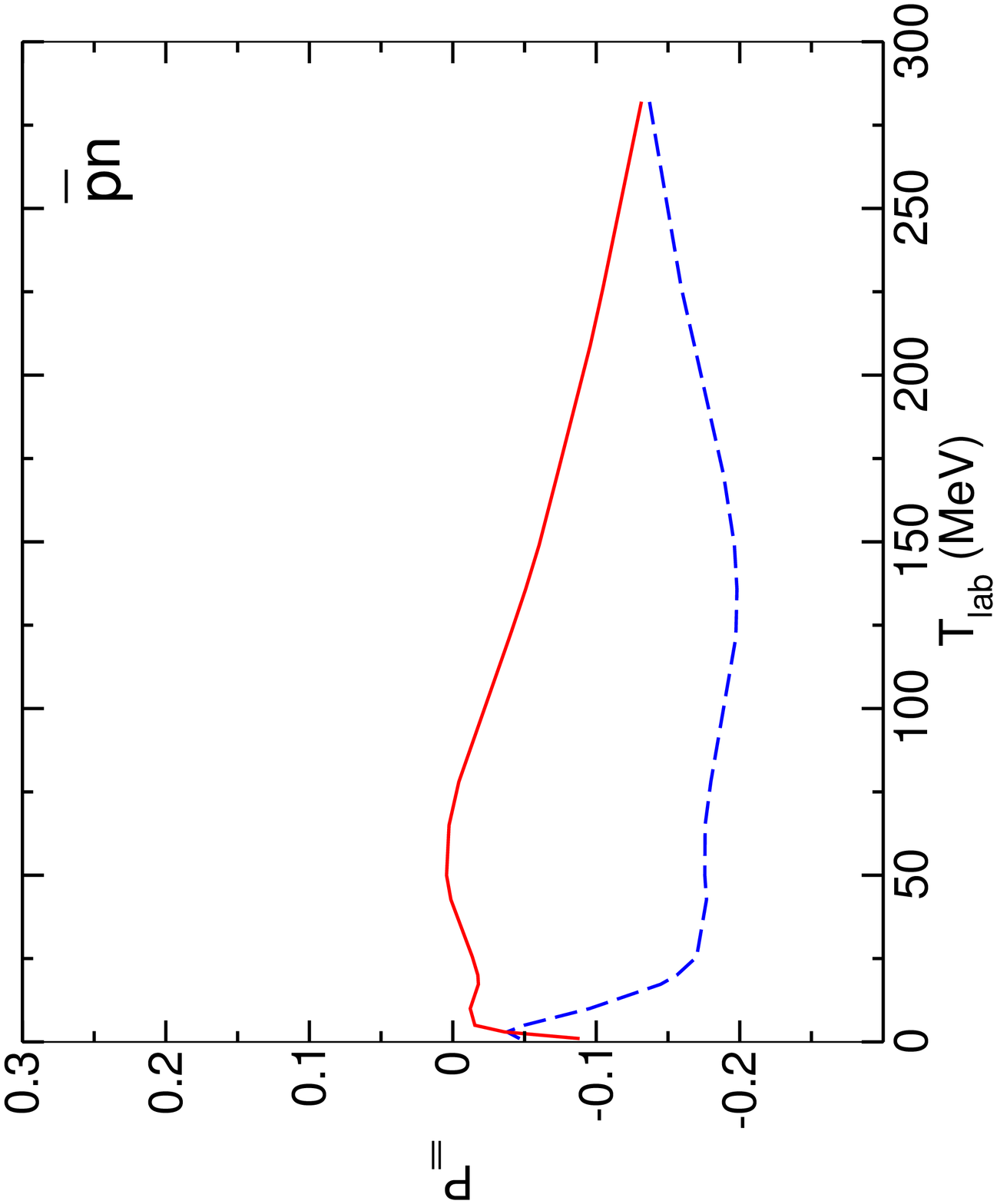}
\includegraphics{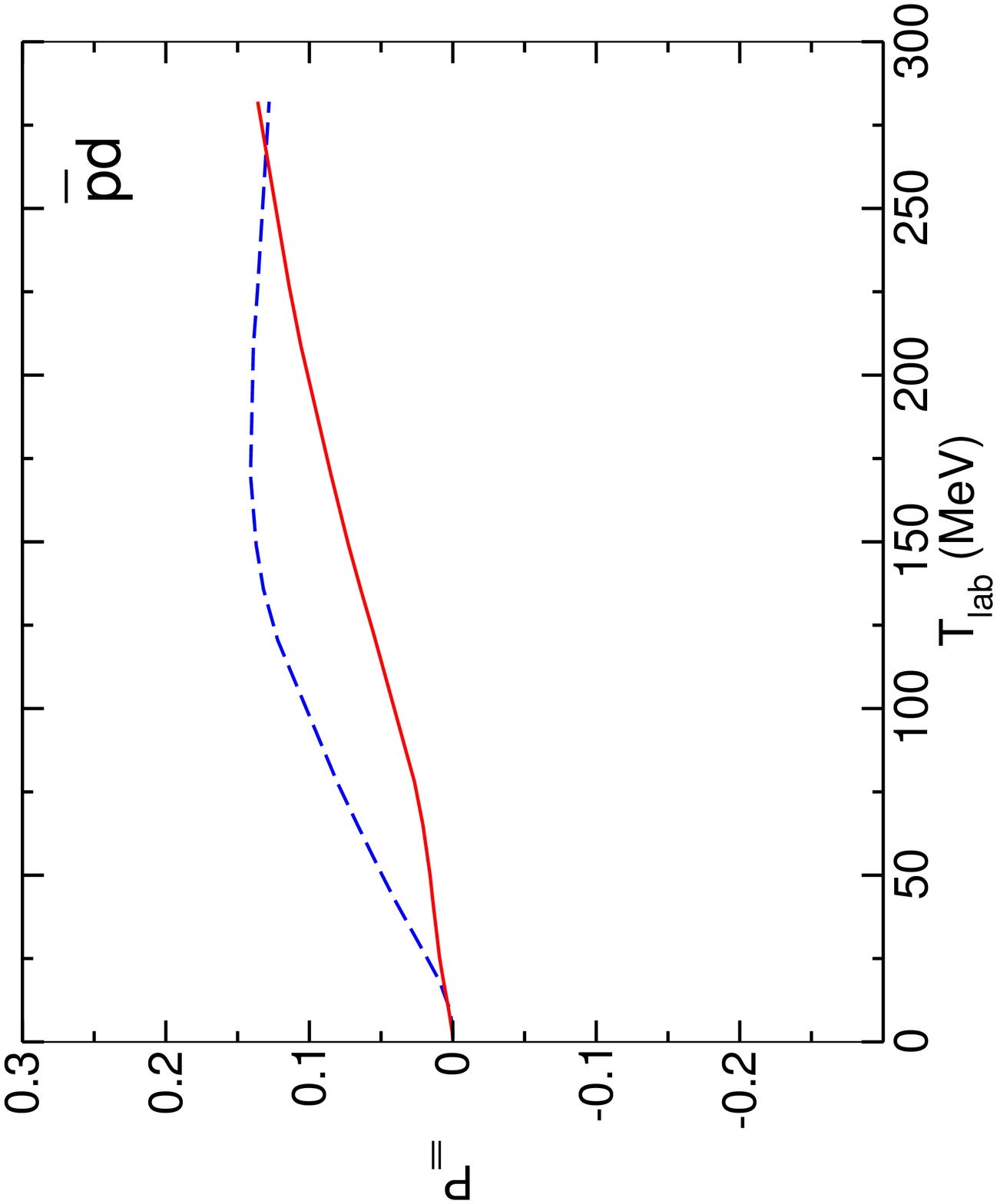}
\vskip 10.5cm
\caption{Dependence of the
 longitudinal polarization $P_{||}$ (i.e. $P_{\bar p}(t_0)$ for
 ${\bfg \zeta}\cdot {\bf m}=1$)
 on the beam energy 
 for the target polarization  $P_T=1$ in the different reactions
 $\bar pp$, $\bar pn$, and $\bar pd$.
The results are for the model A (dashed line) and D (solid line).
The acceptance angle in the cms is $\theta_{acc}=10$ mrad.
}
\label{poldeg2}
\end{figure}

\begin{figure}
\includegraphics{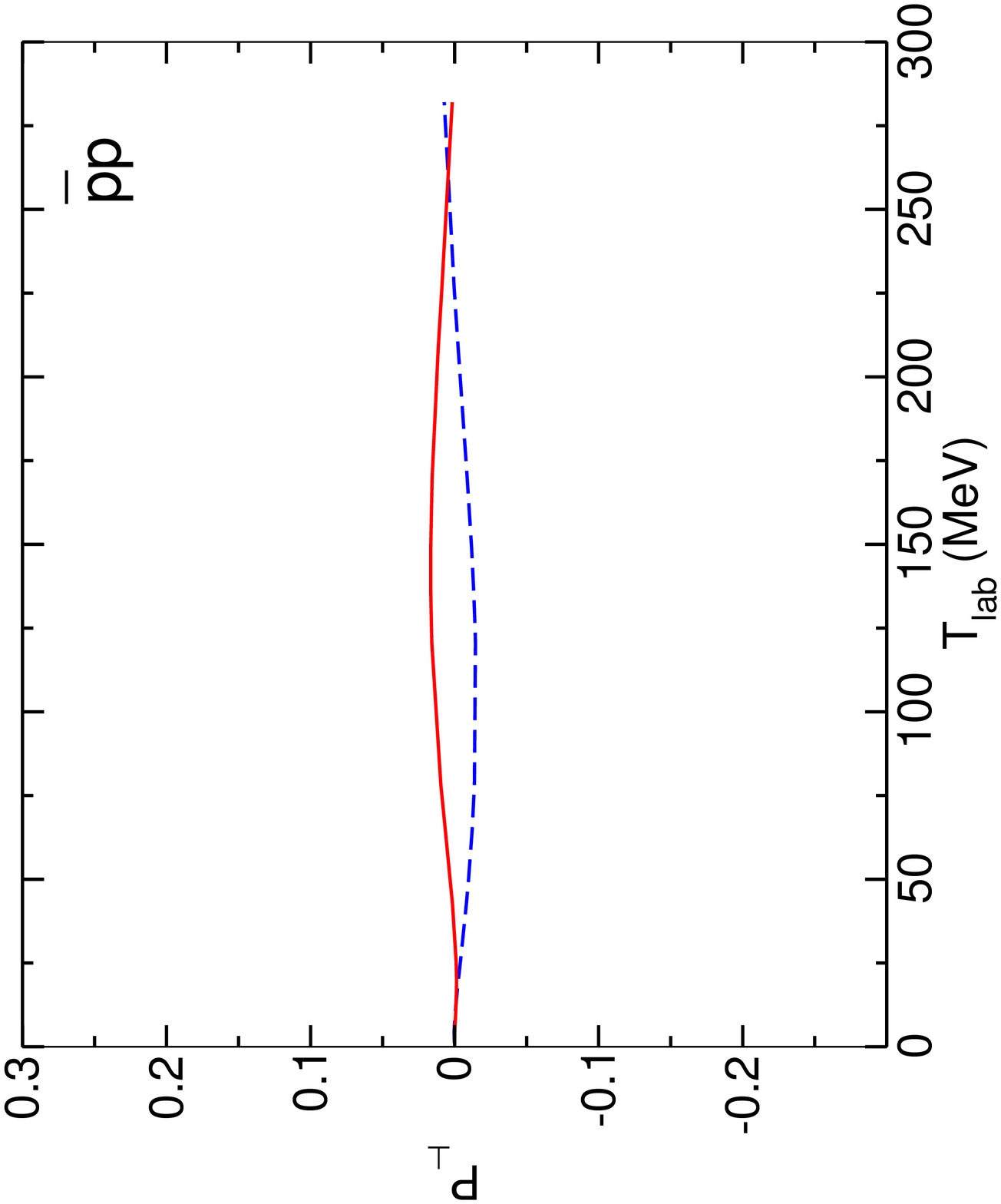}
\includegraphics{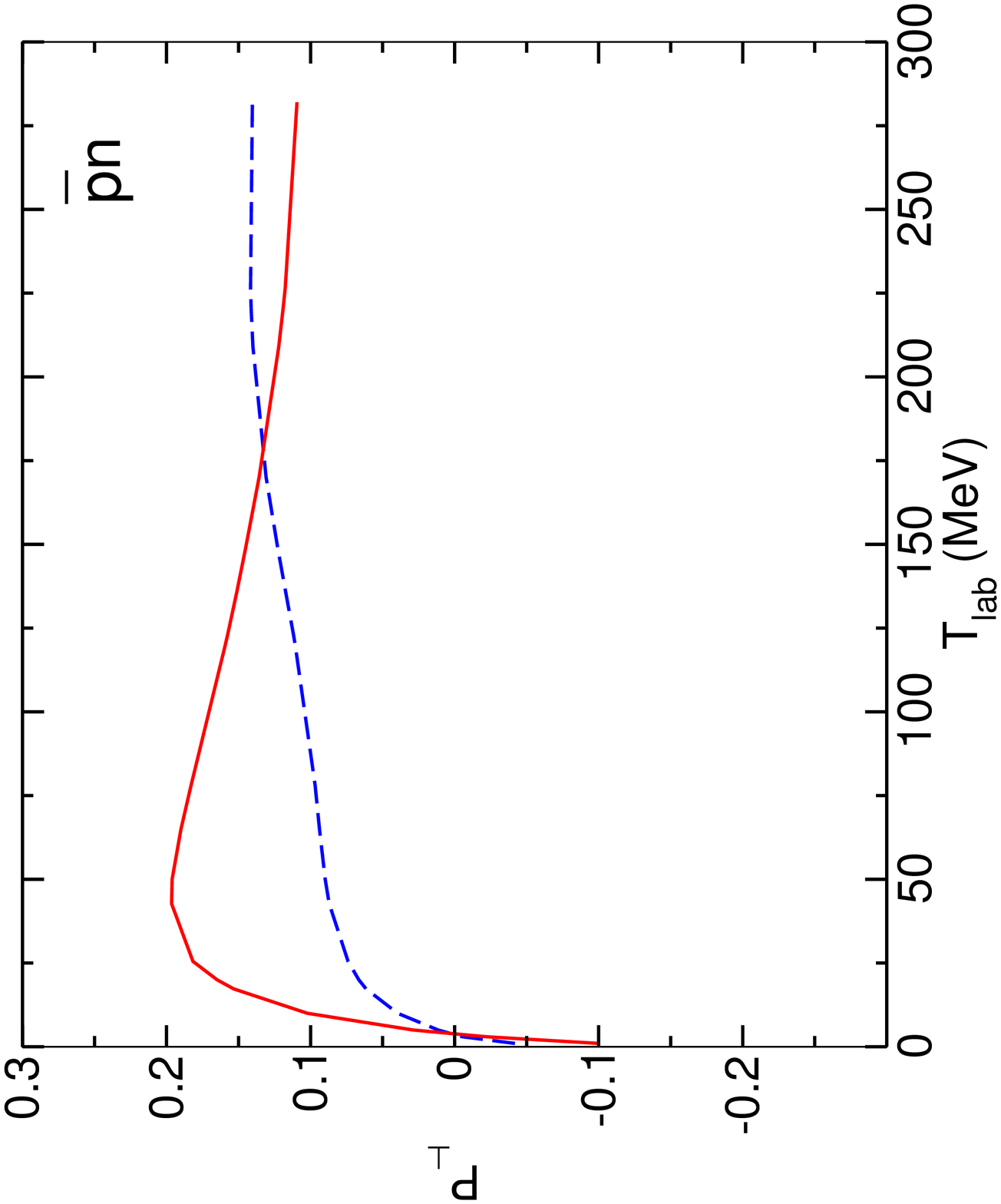}
\includegraphics{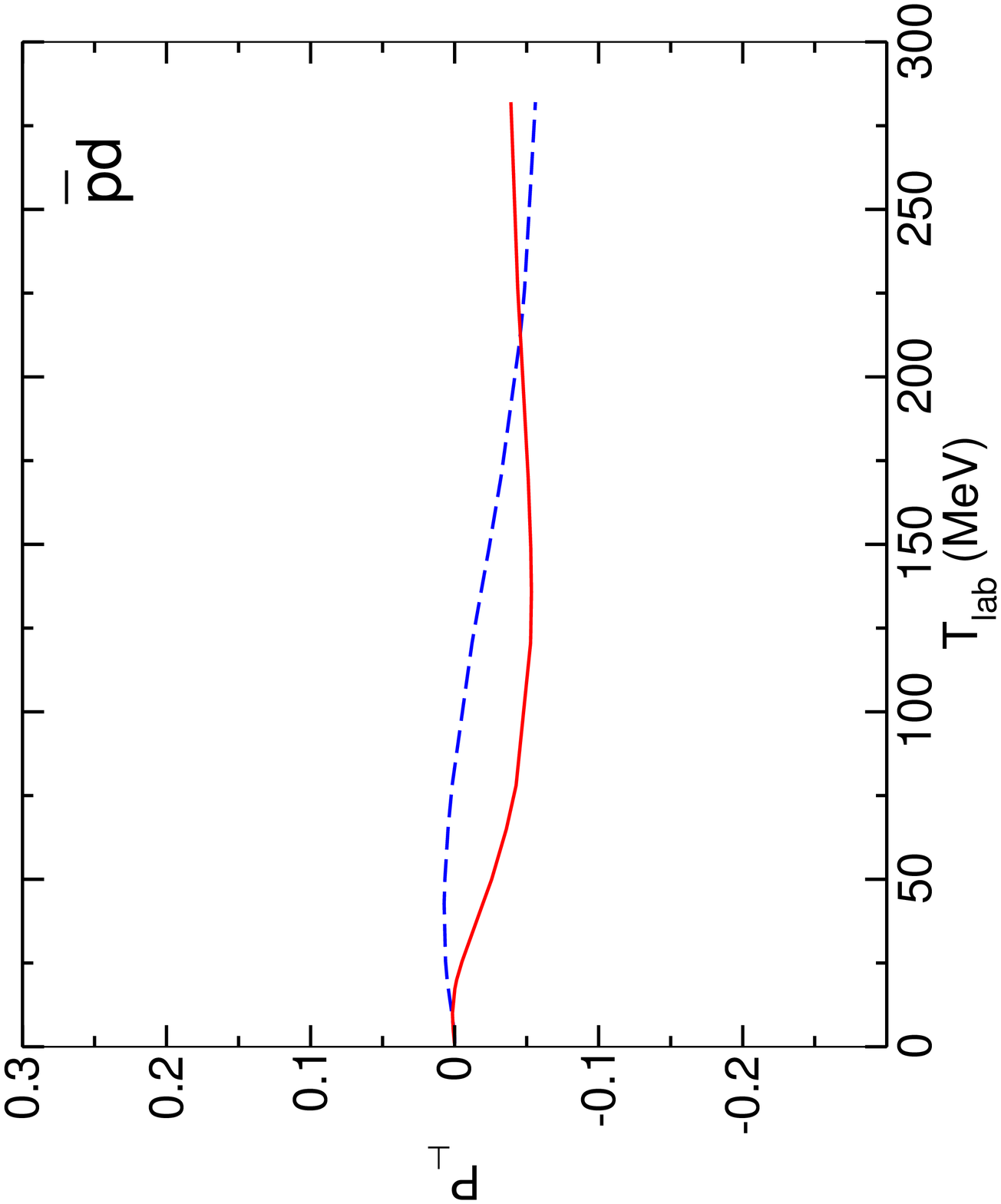}
\vskip 10.5cm
\caption{Dependence of the
 transversal polarization $P_{\perp}$ (i.e. $P_{\bar p}(t_0)$ for
 ${\bfg \zeta}\cdot {\bf m}=0$)
 on the beam energy 
 for the target polarization  $P_T=1$ in the different reactions
 $\bar pp$, $\bar pn$, and $\bar pd$.
The results are for the model A (dashed line) and D (solid line).
The acceptance angle in the cms is $\theta_{acc}=10$ mrad.
}
\label{poldeg1}
\end{figure}

The polarization degree $P_{\bar p}(t_0)$ for   ${\bfg \zeta}\cdot
{\bf m}=1$ ($P_{||}$) at $P_T=P^d=1$
is shown in Fig.~\ref{poldeg2} versus the beam energy.
For the ease of comparison the polarization degree for the 
$\bar p p$ and $\bar p n $ cases are shown too. The results for
${\bfg \zeta}\cdot {\bf m}=0$ ($P_\perp$) are shown in 
Fig.~\ref{poldeg1}.


One can see that, except for the $\bar pn$ case, at energies below 100 MeV
the polarization degree is small due to large total Coulomb cross section.
However, $P_{\bar p}(t_0)$ increases with increasing energy.
For longitudinal polarization maximal values of about 10-15\% are predicted
above 150 MeV. The transversal polarization degree is smaller than the
longitudinal one for both models A and D. For the $\bar p d$ case the
transversal polarization is expected to be larger than for $\bar p p$,
having a maximum of around $5$\%  at 150 -250 MeV (see Fig.~\ref{poldeg1}).
The obtained values for the polarization degree are somewhat smaller
than those presented in \cite{DMS2}, based on the amplitudes of the Nijmegen
$\bar NN$ analysis \cite{Timmermans}.
Experiments for determining the spin-dependent part of the cross sections
of the $\bar pp $ and $\bar p d$ scattering are planned for the near
future \cite{AD,AD1}. Such data should allow one to discriminate between
the different $\bar NN$ amplitudes \cite{DMS2,own}.

\section{Conclusion}

In this work we have used two $\bar NN$ potential models developed by
the J\"ulich group for a calculation of $\bar p d$ scattering within the
 Glauber theory and found that this approach allows one to
 describe the experimental information on (unpolarized) differential and total
 $\bar p d$ cross sections, available at $T_{lab} = 50-180$ MeV, quantitatively.
 For those spin-independent observables the difference in the
 predictions based on those two models turned out to be rather small.

The double-scattering corrections to the unpolarized cross section
were found to be in the order of 15\% in the energy range where the
data are available. But we found that even at such low energies
as 10-25 MeV they are not larger than 20-25\%.
This means that, most likely, the Glauber approximation does work
reasonably well for $\bar p d$ scattering down to fairly small
energies.

The predictions for the spin-dependent cross sections
for $\bar p d$ scattering,
presented in this work, exhibit a fairly strong model dependence, which is due
to uncertainties in the spin dependence of the elementary
$\bar p p$ and $\bar p n$ interactions. Still, for both considered models
we find that the magnitude of the spin-dependent cross sections is
comparable or even larger than those for $\bar p p$.
Thus, our results suggest that $\bar p d$ elastic scattering can be used for
the polarization buildup of antiprotons at beam energies of 100-300 MeV
with similar and possibly even higher efficiency than $\bar p p$ scattering.
However, it is obvious, that only concrete experimental data on the
spin-dependent part of the cross sections of $\bar pp $ and $\bar p d$ scattering
will allow one to confirm or disprove the feasibility of the spin
filtering mechanism for the polarization buildup.

\subsection*{Acknowledgements}
 This work was supported in part by the Heisenberg-Landau program.


\section*{References}
\begin{thebibliography}{99}
\bibitem{PAX} Barone~V  {\it et al}  [PAX Collaboration] 2005 arXiv:0505054[hep-ex]
%
\bibitem{Rathmann}
  Rathmann~F {\it et al}
 2005 {\it  Phys.\ Rev.\ Lett. }   {\bf 94} 014801 
%
\bibitem{FILTEX} Rathmann~F {\it et al} 1993 {\it  Phys. Rev. Lett.} {\bf 71} 1379 
%
\bibitem{MS} Milstein~A I and Strakhovenko~V M 2005 {\it Phys.\ Rev.} E {\bf 72}
 066503
%
\bibitem{NNNP} Nikolaev~N N and Pavlov~F 2005 arXiv:0512051[hep-ph]
\bibitem{NNNP1}
  Nikolaev~N N and Pavlov~F
  2007 {\it AIP Conf.\ Proc.}  {\bf 915} 932

%
\bibitem{HOM} Meyer~H O 1994 {\it  Phys.\ Rev. } E {\bf 50} 1485
%
\bibitem{AD} Lenisa~P and Rathmann~F  [PAX Collaboration] 2005
  arXiv:0512021[nucl-ex].
\bibitem{AD1}
  Barschel~C {\it et al}  [PAX Collaboration] 2009
  arXiv:0904.2325[nucl-ex] 
%
\bibitem{DmitrievMS}
  Dmitriev~V~F, Milstein~A~I and Strakhovenko~V~M 2008
 {\it  Nucl.\ Instrum.\ Meth.}  B {\bf 266} 1122 

\bibitem{own}
  Uzikov~Yu~N and  Haidenbauer~J 2009 {\it  Phys. Rev.} C {\bf 79} 024617

\bibitem{mine}
  Haidenbauer~J,  contribution to this conference. 

\bibitem{DMS2}  Dmitriev~V~F, Milstein~A~I and Salnikov~S~G 2010
  {\it Phys.\ Lett.}  B {\bf 690} 427;
  Salnikov~S~G, contribution to this conference.
%
\bibitem{Hippchen}
          Hippchen~T,  Haidenbauer~J,  Holinde~K and  Mull~V 1991
        {\it  Phys. Rev. C} {\bf 44} 1323;
        Mull~V, Haidenbauer~J, Hippchen~T and  Holinde K 1991
        {\it Phys. Rev.} C {\bf 44} 1337 
\bibitem{Mull}
         Mull~V and Holinde~K 1995 {\it Phys. Rev. C} {\bf 51} 2360

\bibitem{rekalo}  Rekalo~M~P, Piskunov~N~M and Sitnik~I~M 1998
 {\it Few-Body Syst.} {\bf 23} 187 

%
\bibitem{FrancoGlauber} Franco~V and Glauber~R~J 1966 {\it  Phys. Rev.} {\bf 142}
1195 
%
\bibitem{Kondratyuksb} Kondratyuk~L~A, Shmatikov~M~Zh and Bizzarri~R 1981
 {\it Yad. Fiz.} {\bf 33} 795  [1981 {\it Sov. J. Nucl. Phys.} {\bf 33} 413 ]
%
\bibitem{Dalkarov} Dalkarov~O~D and Karmanov~V~A 1985
{\it Nucl.\ Phys.}  A {\bf 445} 579

%
\bibitem{BizzarriNC74} Bizzarri~R {\it et al} 1974 {\it  Nuovo Cim.} {\bf 22 A} 225 
%
\bibitem{Kalogeropoulos}  Kalogeropoulos~T and Tzankos~G~S 1980 {\it Phys. Rev.} D
{\bf 22} 2585 
%
\bibitem{Burrows} Burrows~R~D {\it et al} 1970 {\it  Austr. J. Phys.}
 {\bf 23} 919 
%
\bibitem{Carroll}  Carroll~A~S {\it et al} 1974 {\it  Phys. Rev. Lett.} {\bf 32}
 247 

\bibitem{Hamilton}  Hamilton~R~P {\it et al} 1980 {\it  Phys. Rev. Lett}. {\bf 44}
 1182 
%
\bibitem{ABB} Aladashvili~B~S {\it et al} 1977 {\it J. Phys.} G {\bf 3} 7 
%
\bibitem{bruge88} Bruge~G {\it et al} 1988 {\it  Phys. Rev.} C {\bf 37 } 1345

\bibitem{Timmermans} Timmermans~R, Rijken~T~A and de Swart~J~J 1994
  {\it Phys.\ Rev.}  C {\bf 50} 48 
%

\end{thebibliography}
\end{document}